# Multimodal Characterization of Emotion within Multimedia Space

Dayo Samuel Banjo*, Connice Trimmingham*, Niloofar Yousefi*, Nitin Agarwal*
COSMOS Research Center, UA - Little Rock, Arkansas, USA
{sbanjo, ctrimmingham, nyousefi, nxagarwal}@ualr.edu

*Abstract*— Technological advancement and its omnipresent connection have pushed humans past the boundaries and limitations of a computer screen, physical state, or geographical location. It has provided a depth of avenues that facilitate human-computer interaction that was once inconceivable such as audio and body language detection. Given the complex modularities of emotions, it becomes vital to study human-computer interaction, as it is the commencement of a thorough understanding of the emotional state of users and, in the context of social networks, the producers of multimodal information. This study first acknowledges the accuracy of classification found within multimodal emotion detection systems compared to unimodal solutions. Second, it explores the characterization of multimedia content produced based on their emotions and the coherence of emotion in different modalities by utilizing deep learning models to classify emotion across different modalities.

## I. Introduction

RECENT history has highlighted many radical social shifts and transformations. From genetic engineering, ending of the Cold War, and most noteworthy the emergence of technology [1]. The rapid development of the information age, specifically social media, has become the greatest offspring of technological revolution due its factor of instant real-time connection [2]. At its inception, textual content was the main source of information over the internet [3]. The invention of the World Wide Web developed a networked communion frame that facilitated expressive blogging platforms and services such as email wikis and weblogs [3]. As social media—and its adjacent space expand across geographical locations, real time information became easily accessible to millions [3]. Multimedia platforms' traffic traction increased exponentially and consumers gravitated to daily video updates, online news, and entertainment [3], [4].

With the maturity of new technologies, there is a new emergence of different social structures, perspectives, data sources and influence [4]. Social media platforms are now one of the main tools used in modern day human machine interaction and offer multimedia data ranging in different formats such as text, audio, images, and videos [5]. This emergence has brought about an auxiliary, but notable factor: the demand and taste for digital content shifted and there was a new magnetism phenomena between multimedia spaces such as YouTube and consumers [6], [7].

Many research has proven inadequacies within a unimodal emotion recognition system to satisfy the demands of the current emotional states found on multimedia platforms [8]. Researchers have shown that emotion information contained in a single mode is limited [8], [9]. With the understanding that a fusion model can be utilized to optimize performance of social media emotional recognition systems, this study would look at the internal connection between text and speech to classify the final emotional state found within videos on YouTube. This study will briefly explore and acknowledge prior studies that proved higher accuracy levels can be achieved when a fusion multimodal model is applied to assess emotion. It would further articulate our anchorage of machine learning techniques on audio and textual content to describe the emotions of content producers. The differing nature of emotions presents a twofold dynamic. It produces a vast body of literature and simultaneously prevents an encyclopedic review [9]. This article will start by reviewing essentials to support this work and discuss the empirical study and findings. We will conclude by exploring limitations, and applications for future work.

## II. Literature Review

### A. Emotions and Uni-Model Approaches

In prior works, various uni-model approaches have been done for emotion detection. Some studies detected the emotion from speech [10], [11] and other studies claim that the instructive way to understand emotions is from facial expressions in images and videos [12]-[14]. Researchers also presented emotion recognition based on audio and voice streams [15], [16]. Hatice Gunes et al., claims that facial expressions can provide useful emotional information. However, sometimes the facial expression is not able to discriminate the emotional states as some face movements are not distinguishable [17]. Recent years' researchers have used multimodal contents mainly including text, audio, and video for emotion detection [18]. By using multiple modalities, we can cover more information rather than single modalities, which will help to improve the robustness of emotion detection [19].

### B. Multi-Model Approaches

Muharram Mansoorizadeh et al., proposed an approach for multimodal feature extraction and combination in which the performance of his approach is higher than the unimodal systems such as the face-based model and speech-based model [20] Chin Xang et al., claim that in published papers, the authors indicated only cases of superiority in multi-modality over audio-only or video-only modality. So, they propose a novel framework to address the Audio-Video Emotion Recognition (AVER) problem and their MRPN solution contributes to a better average recognition rate of approximately 2% [21]. Valentin Gabeur et al., propose a novel framework that is a multimodal transformer to jointly encode the different modalities in the video. They use their model on the various datasets and compare them with different methods and obtain better Recall results for video retrieval [22]. David Harwath et al., in their research explore neural network models that learn to associate segments of audio captions with the relevant portions of natural images that they refer to, by embedding audio segments and images in the same space without the need for annotated training data in either modality [23]. Yao-Hung Hubert Tsai et al., discussed two major challenges in modeling multimodal human language time-series data, and they introduce the Multimodal Transformer (MulT) to address the issues in an end-to-end way. The results obtained from MulT for multimodal emotions analysis (aligned and non-aligned multimodal sequences) show higher accuracy and F1-score than other previous methods [24].

### C. Transformers

Transformer is a novel deep learning architecture which has been successful in many fields of artificial intelligence such as natural language processing, computer vision, and audio processing. Transformer-based pre-trained models can perform state-of-the-art on a variety of tasks. [25]. It is a neural network architecture used in sequence-to-sequence tasks that adopts a self-attention mechanism to compute representations of its input and output. It enhances contextual information understanding in long dependencies of sequence data, using the attention mechanism. Much research has contributed to the ongoing development of the transformer model. Tay et al. characterized a large body of the "X-former" model, aimed at providing a comprehensive overview of existing work across multiple domains [26]. Researchers have applied transformer architecture to assess historical release of underground water contaminants, achieving more accuracy in analyzing water distributions and planning [27]. Given its high level of effectiveness across multiple domains such as language and vision, researchers have leveraged transformer models in sentiment analysis to assess drug reviews and improve pharmacovigilance systems [28].

### D. Self-attention

Self-attention is an attention mechanism calculating a representation of a sequence considering the temporal relationship of all items in the sequence. Self-attention transformers have been significantly used in several domains of natural language processing, speech recognition, and computer vision. Zhang et al., used the self-attention transformer model for speech recognition and compared the results with a vanilla transformer [29]. Berg et al., introduced a keyword transformer using a fully self-attention transformer [30]. Vittorio et al., developed a self-attention transformer model for human action recognition [31]. Han et al., introduced a model for speaker verification using the local information model and self-attention transformer [32]. Mehboob et al., developed a self-attention transformer model to diagnose the COVID-19 using Computed Tomography (CT) imaging technique and compared the results with CNN and Ensemble classifiers [33].

### E. Bert Transformer

Bidirectional Encoder Representations from Transformers (BERT) is a machine learning method for NLP. BERT is a deep learning pre-trained model which is used to pre-train deep bidirectional representations from a text [34]. Francisca et al., in their paper compares the BERT model and other transformers-based models and discusses the pros and cons for them. They mentioned the strength of the BERT in NLP tasks for emotion detection from texts [35]. Acheampong et al., in their research to analyze the efficiency of using transformer encoders for emotion detection, propose a framework using BERT for the first stage and Bi-LSTM for the second stage [36]. Andrea et al., applied sentiment analysis and emotion detection by using the BERT pre-trained language model on Twitter data and evaluating the performance of the model [37].

### F. VisualBert

In order to classify multimodal data (text and audio), we fine-tune VisualBERT on the Interactive Emotional Dyadic Motion Capture (IEMOCAP) dataset. VisualBERT is a transformer model that leverages a simple and flexible framework for modeling a broad range of vision-and-language tasks [36]. Li et al. proposed VisualBERT as a stack of transformer layers with self-attention [36]. The advantages of incorporating VisualBERT and another multimodal transformer variant for joint tasks involving vision and natural language has shown high accuracy in performance. VisualBERT transformer network implicitly aligns textual and visual elements by re-using the self-attention mechanism within the transformer. It provides a contextualized representation of visual and textual elements.

## III. METHODOLOGY

This section provides an overview of our study, data collection perspective, and approaches used to characterize emotion within the multimedia space. It focalizes on the coherence characterization of audio and text modality along with the relative performance comparison of unimodal and multimodal approaches for emotion characterization.

### A. Data Collection

Given the complexities of emotions, we discuss the unimodal and multimodal characterization of emotion from multimedia content using a variant of the transformer model for our study. The first process was the collection of data, in which we employed the use of YouTube API for videos' and metadata's data to generate the YouTube 1K database.

This task was conducted on two datasets, YouTube 1K dataset, and IEMOCAP [36] database.

1) **YouTube-1K Dataset** consists of 1000 video IDs, selected from an arbitrarily 30 channels from eight categories, namely games, travel and events, sports, entertainment, science and technology, news, and politics. It contains multimedia information - audio, frames, title, and descriptions. We perform unimodal emotion analysis on the audio and textual (title and description) information extracted from this data with YouTube API, using pre-trained models to evaluate the unimodal coherence of emotion. Fig. 1 shows the number of videos for each category.

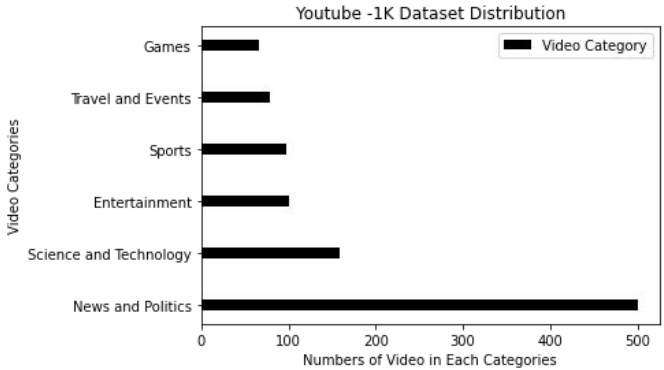

Fig. 1 Top Level Category Statistics of the YouTube 1K Dataset

2) **IEMOCAP** The Interactive Emotional Dyadic Motion Capture (IEMOCAP) is a widely used dataset in multimodal emotion assessment. It contains 1440 multimedia files, with each file represented in audio, visual, and text transcription from acted, multispeaker, and multimodal databases. The IEMOCAP database has an annotation for multiple categories such as neutrality, sadness, anger, surprise, fear, and happiness.

### a. Unimodal Coherence Analysis

To study inter-signal coherence between unimodal emotion analysis on the YouTube data extracted, we applied an emotion characterization framework. Our process consisted of using a pre-trained Transformer text-based emotion model DistilBERT-base-uncased-emotion [39] and audio-based emotion characterization using librosa [40], a python package that analyzes audio sound effectively. Given a set of $N$ videos $\{v_i, ..., v_n\}$ we obtain two sets of audio and textual data: $\{t_i, ..., t_n\}$ from videos, description and title, $\{a_i, ..., a_n\}$ from videos' audio. The audio and video were process separately to produce a set of results described as: $E(x) = E^{m \times n}$ where, $x$ is the modality being considered and can be audio or text, $m$ is the number of classes and $n$ is the number of videos. The result output for each audio signal or text signal, contains probabilities for each class. Emotion labels for the classes are anger, fear, happy, love, sad and surprise.

### b. Multimodal Single Shot Analysis

The goal of this analysis is to compare multimodal emotion assessment with unimodal solutions on the same dataset. For this purpose, we fine-tuned and pre-trained VisualBERT [38] with a combinatorial input of text and audio modalities for the multimodal analysis on the IEMOCAP and also ran the dataset with the librosa library for the comparison with unimodal. The combinatorial input for the multimodal was a spectrogram image of the audio wave signal and a transcribed text representation of the audio signal. VisualBERT is a simple vision and language model consisting of a stack of Transformer layers that implicitly align an element of input text and regions in an associated input image with self-attention. This model was chosen due to it contextual inference of different modalities in a single pass; we fine-tuned VisualBERT on an IEMOCAP test set which shows high accuracy in detecting dependency between the relationship with direct supervision and can provide a joint model for visual and textual representation learning, to predict numeric emotion score for six classes of emotions. The model was fine-tuned by replacing the last output layer of the VisaulBERT with a dense layer and a sigmoid function. The pipeline of the fined model is constructed with the illustration in Fig. 2.

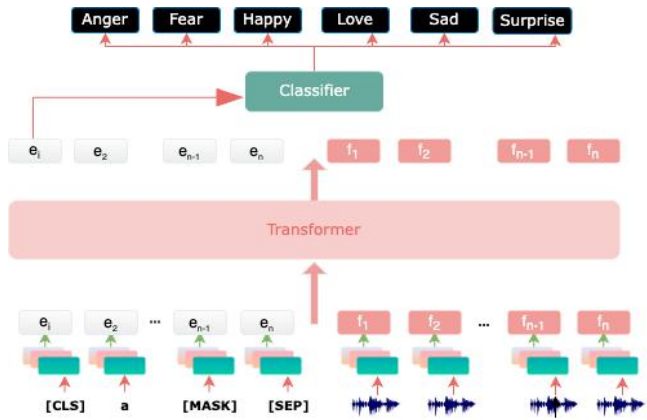

Fig. 2 Fined Tuned VisualBERT Architecture

## IV. EVALUATION AND RESULTS

The quantitative result is divided into two parts; the first part is to observe the coherence between different emotion modality analyses on the YouTube-1K dataset, and the second part is to compare unimodality with multimodality using the IEMOCAP dataset.

### B. Unimodal Coherence Evaluation

To further investigate the coherence and the disparity between the Text-Based unimodal Assessment result distribution and the Audio-Based unimodal Assessment distribution, we implore a few statistical methods. Statistical methods [41] provide an inherent quantitative measure that effectively analyzes the dimensions' features, such as the central tendency, the measure of dispersion, and distribution type. The following are the metrics used for our quantitative analysis: Cross-Correlation and Measures of Central Tendency.

#### 1) *Cross Correlation of Unimodal*

Emotions emitting in a video can take many modularities, thus to accurately determine the correlation of the two different emotions within our corpus of data, we calculate the correlation coefficient of the unimodal audio emotion assessment distribution and the unimodal text emotion assessment. The resultant correlation coefficient was to understand how different modality time series of how for each emotion in a different modality, we see a negative or positive correlation. The result of the correlation was shown in Table 1; this shows there is a positive correlation in the happy, anger, fear and love emotions for both modality, while the remaining signals gave a weak coherence. This analysis suggests a high probability that our selected videos were focused on subject matter that invoked happy, anger, fear, love emotions but no correlation was observed for sad and surprise emotions.

TABLE I
CORRELATION COEFFICIENT FOR AUDIO AND TEXT UNIMODAL COHERENCE

| Emotion | Coefficient |
|---|---|
| Anger | 0.805 |
| Fear | 0.781 |
| Happy | 0.720 |
| Love | 0.950 |
| Sad | 0.054 |
| Surprise | 0.007 |

Fig. 3 shows the correlation plot of audio emotion scores against text emotion score for different emotion signals. Similarly, from the plot we can infer that anger, fear, happy, and love emotions show similar correlations across various modalities of emotion analysis compared to sad and surprise emotions.

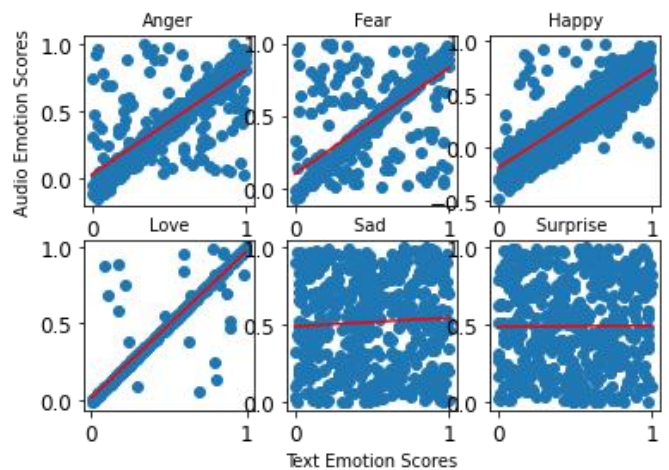

Fig. 3 Emotion Correlation Coefficient

#### 2) *Measure of Central Tendency*

To understand the pattern distribution between audio and text modality, we use two measures of central tendency techniques to learn the signature of the distribution, as shown in Fig. 4 for each label emotion, mean and median distributions. The mean result captures the center value of the emotion; the median distribution provides empirical evidence for outliers in the dataset. The mode shows the common values in each emotion for different modalities. The range in comparison with the mean median and mode shows whether there are outliers, skew, or imbalance distribution; with these values, we can say that the prediction of the audio emotion is more concise because we see less outliers. We can determine if there is an outlier in the distribution through the range.

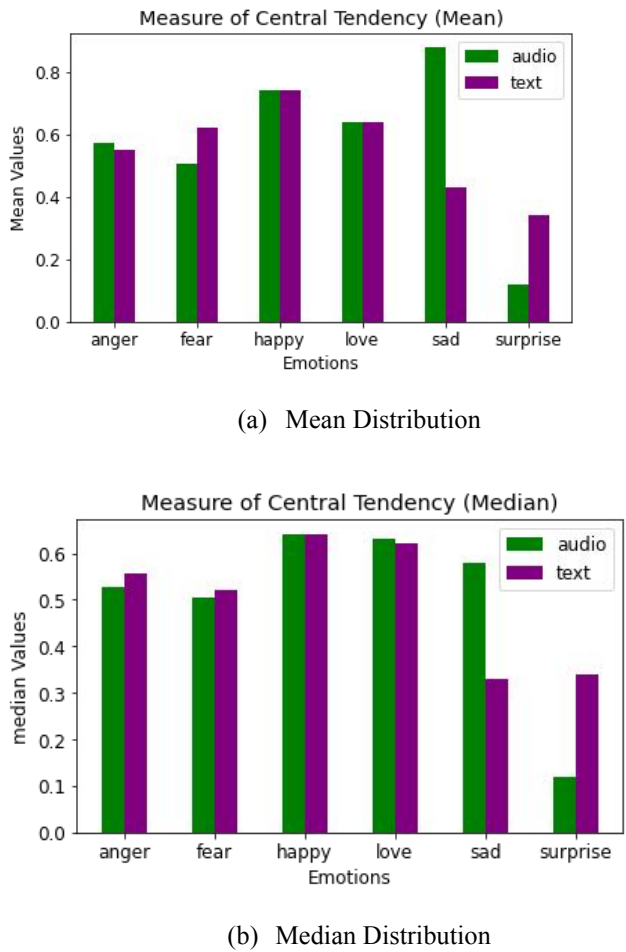

(a) Mean Distribution

(b) Median Distribution

Fig. 4 Mean and Median Distribution of Audio and Text Measure of Central Tendency.

### 3) Comparison of Unimodal and Multimodal Analysis

To investigate the performance of unimodal and multimodal analysis, we compute the validation and test accuracy, true positive, true negative, false positive and false negative of each emotion for each model. The validation and accuracy results of the models on the IEMOCAP dataset compared to validation accuracy and test accuracy as shown in Table 2, prove that multimodal (Audio + Text) performs better than unimodal audio only on emotion assessment; thus, the multimodal assessment achieves higher accuracy.

TABLE 2
DIFFERENT MODALITY ACCURACY COMPARISON

| Modality | Validation Accuracy (%) | Test Accuracy (%) |
|---|---|---|
| Unimodal | 72 | 71.6 |
| Multimodal | 74.6 | 73.5 |

To further analyze the accuracy of our result, we plot the confusion matrices of both the unimodal and multimodal emotions classification result. This process allowed a thorough examination of the predicted values against the actual classes of both unimodal and multimodal models shown in Fig. 5. Our observation shows the greater potential result from multimodal analysis in predicting the emotions than unimodal emotions.

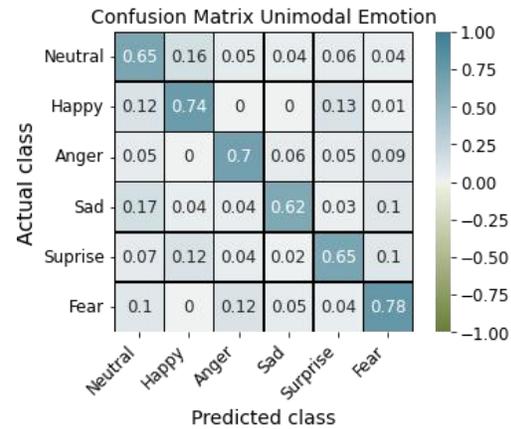

a. (a) Unimodal Confusion Matrix

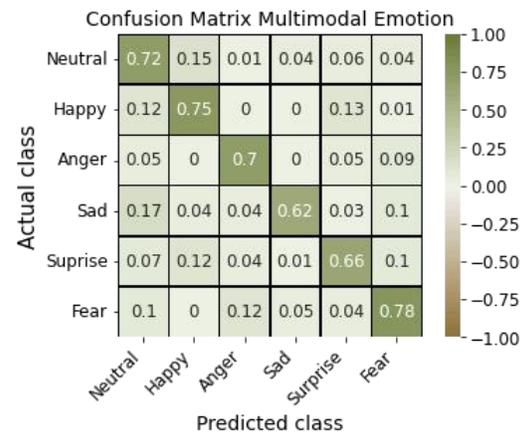

(b) Multimodal Confusion Matrix

Fig. 5 Unimodal and Multimodal Confusion Matrices for Six Emotions Classes

Fig. 6 shows the AUROC (Area Under the Receiver Operating Characteristic) in the validation set and test set for unimodal and multimodal emotion assessment on the IEMOCAP dataset. Multimodal has a higher AUC showing good measure of emotions separability. This highly suggests that multimodal emotion outperform unimodal in distinguishing the emotion classes exerted in the data.

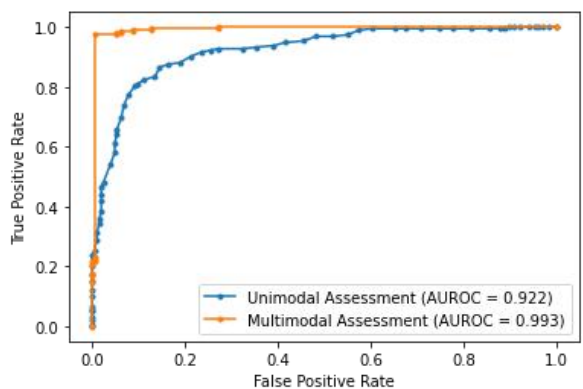

Fig. 6 Area Under the Curve Region Operating Characteristics Curve (AUROC) for Unimodal and Multimodal Assessment.

## V. CONCLUSION

We proposed a multimodal single shot multimedia content emotion analysis to assess how information producers push emotions based on titles, descriptions, and audio waveform, and compare multimodal emotion assessment with unimodal solutions. Our analysis shows that videos emitting emotions of happiness, anger, fear and love were proportionally coherent within both text and audio signals. Furthermore, we fused the result from the multimodal and unimodal analysis to generate results for classification. We found high performance in our multimodal fusion architecture when compared to a unified model. This study helps to understand that a multimodal approach can be utilized to optimize the performance of social media emotional recognition systems; more so, characterize emotion based on coherence and disparity of different modality results. Future research can explore the dimension of emotion assessment incorporating more advanced multimodal features simultaneously, such as video, audio, image and text.

## ACKNOWLEDGMENT

This research is funded in part by the U.S. National Science Foundation (OIA-1946391, OIA-1920920, IIS-1636933, ACI-1429160, and IIS-1110868), U.S. Office of Naval Research (N00014-10-1-0091, N00014-14-1-0489, N00014-15-P-1187, N00014-16-1-2016, N00014-16-1-2412, N00014-17-1-2675, N00014-17-1-2605, N68335-19-C-0359, N00014-19-1-2336, N68335-20-C-0540, N00014-21-1-2121, N00014-21-1-2765, N00014-22-1-2318), U.S. Air Force Research (FA9550-22-1-0332), U.S. Army Research Office (W911NF-20-1-0262, W911NF-16-1-0189), U.S. Defense Advanced Research Projects Agency (W31P4Q-17-C-0059), Arkansas Research Alliance, the Jerry L. Maulden/Entergy Endowment at the University of Arkansas at Little Rock, and the Australian Department of Defense Strategic Policy Grants Program (SPGP) (award number: 2020-106-094). Any opinions, findings, and conclusions or recommendations expressed in this material are those of the authors and do not necessarily reflect the views of the funding organizations. The researchers gratefully acknowledge the support.